\documentclass[11pt]{article}
\usepackage[colorlinks=true, citecolor=blue, urlcolor=blue, linkcolor=blue, breaklinks=true, pdfpagelabels=false]{hyperref}
\usepackage{cite}

\usepackage{hyperref}

\usepackage{amsmath,amsfonts,amssymb}
\usepackage{graphicx}
\usepackage{epsfig,epsf}
\usepackage{bm}
\usepackage{xcolor}

\def\be {\begin{equation}}
\def\ee {\end{equation}}
\def\bea {\begin{eqnarray}}
\def\eea {\end{eqnarray}}

\def\barr{\begin{array}}
\def\earr{\end{array}}

\def\opcit(#1){ {\em op. cit.}, #1}

\def\issue(#1,#2,#3){#1, #2 (#3)} 

\def\equationautorefname~#1\null{Eq.\,(#1)\null}
\def\pageautorefname\nobreakspace{p.}

\makeatletter\renewcommand{\p@subsection}{\thesection.}\makeatother%
 
\begin{document}

\renewcommand*{\thefootnote}{\fnsymbol{footnote}}


\begin{center}
{\Large\bf{Stress testing of fast reconstruction pipelines using machine learning}}


\vspace{5mm}

{\bf Swagata Ghosh}$^{a}$\footnote{swgtghsh54@gmail.com}

\vspace{3mm}
{\em{${}^a$\ \ \ Central University of South Bihar, Gaya, Bihar, India.
}}

\end{center}

\begin{abstract}

The fast reconstruction and detector-simulation pipelines are widely used in different scientific domains, such as, in High Energy Physics (HEP) and medical imaging, where the full experimental or the device-level simulation is computationally challenging. 
Instead of the use of the global data context, these pipelines use simplified response models which assume reconstruction uncertainty relies on local input parameters. 
To probe the robustness of the local assumption, a domain-agnostic context-aware stress testing pipeline is introduced, and a reconstruction response depending on the global parameters is also allowed. 
For an instance in HEP at High-Luminosity LHC (HL-LHC) simulation, this work shows that the decay channel $Z \rightarrow \ell\ell$, as a benchmark, violates the local assumption resulting in a significant reconstruction bias and resolution degradation. 
Using the unsupervised regime-mapping framework, this work also restores this peak stability and recovers the truth-level resolution, where a robust diagnostic tool for next-generation fast simulation pipelines is accommodated. 

\end{abstract}



\setcounter{footnote}{0}
\renewcommand*{\thefootnote}{\arabic{footnote}}

\section{Introduction}
\label{intro}

In any scientific discovery, including High Energy Physics (HEP) and Medical Imaging (MI), precise reconstruction of the observables is a fundamental requirement. 
In HEP, probing the Standard Model (SM) as well as searching for new particles demand the precise reconstruction of particle masses and decay kinematics. 
The standard reconstruction pipelines rely on the uniform distribution of data across the entire phase space. 
However, the work of these references \cite{Mankel:2004yv,CMS:2026znb,CMS:2024irj} shows that the reconstruction accuracy is challenged frequently in the non-bulk regions due to this assumption. 
The dilepton invariant mass $(m_{\ell\ell})$ in this paper also evidences this limit -- 
the reconstruction accuracy is challenged in the tail of the distribution as the bulk of data are accumulated in the $Z$-peak region, and in high energy regimes the particles are boosted. 

To address the robustness of the mass reconstruction pipeline, this work employs an unsupervised, lightweight Machine Learning (ML) framework. 
The ``unsupervised'' nature of this ML framework allows its ability to probe the data-driven patterns without the bias of the already existing supervised labels \cite{Bardhan:2024zla,Andreassen:2018apy}. 
In this work, unsupervised clustering is used to characterize the global kinematic structure, instead of depending only on the local input parameters, and we have, as a result, a map of identified context regime where data are partitioned on the basis of the scalar sum of the lepton transverse momenta ($H_T$) and the absolute pseudorapidity separation ($|\Delta\eta|$). 

Based on these classified regimes, one can examine the performance of the reconstruction pipeline across these regions. 
This framework points out the stability in the bulk regions, whereas the stress tests exhibit the divergence in performance between regimes. 
Specifically, the resolution degrades substantially in the high $H_T$ regimes. 
Moreover, the heat map analysis illustrates that this inaccuracy correlates with increasing $H_T$, and for a traditional pipeline a danger zone exists. 

This approach provides a concrete roadmap to quantify performance for targeted, context-sensitive regimes, instead of a uniform global average. 
In HEP, this work provides a path towards an improving overall framework robustness and precision for future High-Luminosity LHC (HL-LHC) physics \cite{HEPSoftwareFoundation:2017ggl}. 
This approach can also be utilized in MI \cite{wang2016perspectivedeepimaging}, but that is beyond the scope of this paper, and hence we leave it for our future work. 

\begin{figure}
 \begin{center}
  \includegraphics[width=5.9cm]{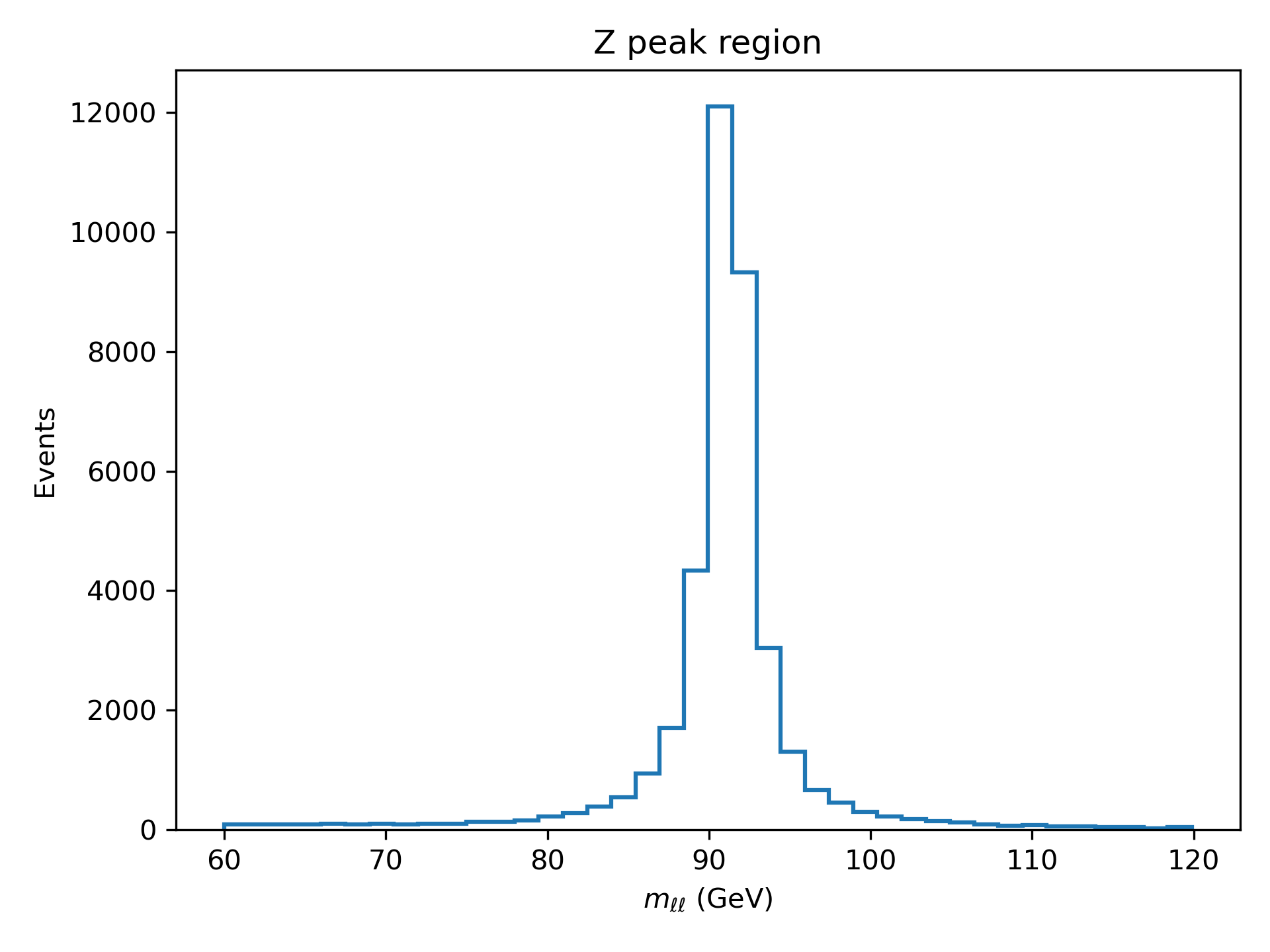} \ \ 
  \includegraphics[width=5.9cm]{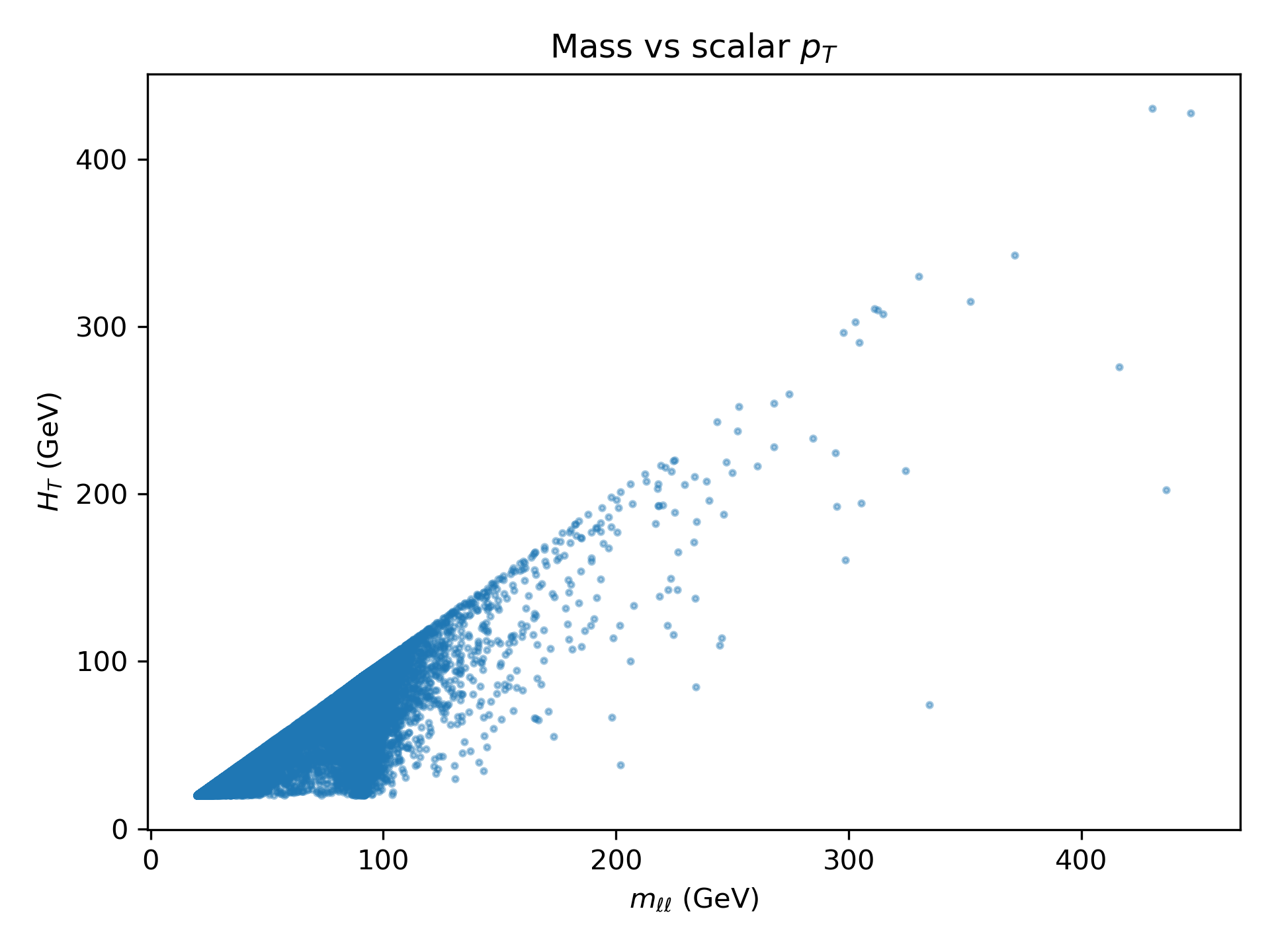}  
 \label{fig:mll}
 \end{center}
\caption{\scriptsize Left : The dilepton invariant mass distribution $m_{\ell\ell}$ within $60 - 120$ GeV reveals the $Z$-boson resonance peak. 
\newline
Right : Scatter plot of invariant mass $m_{\ell\ell}$ as a function of the global context scale $H_T$.}
\end{figure}

In this paper, section \ref{Analysis} analyses the way to execute the results provided in section \ref{Results}. 
Finally, section \ref{Conclusion} concludes with a note on the utilization of this work in Medical Imaging in our future work.

\section{Analysis}
\label{Analysis}

\begin{figure}
 \begin{center}
  \includegraphics[width=5.9cm]{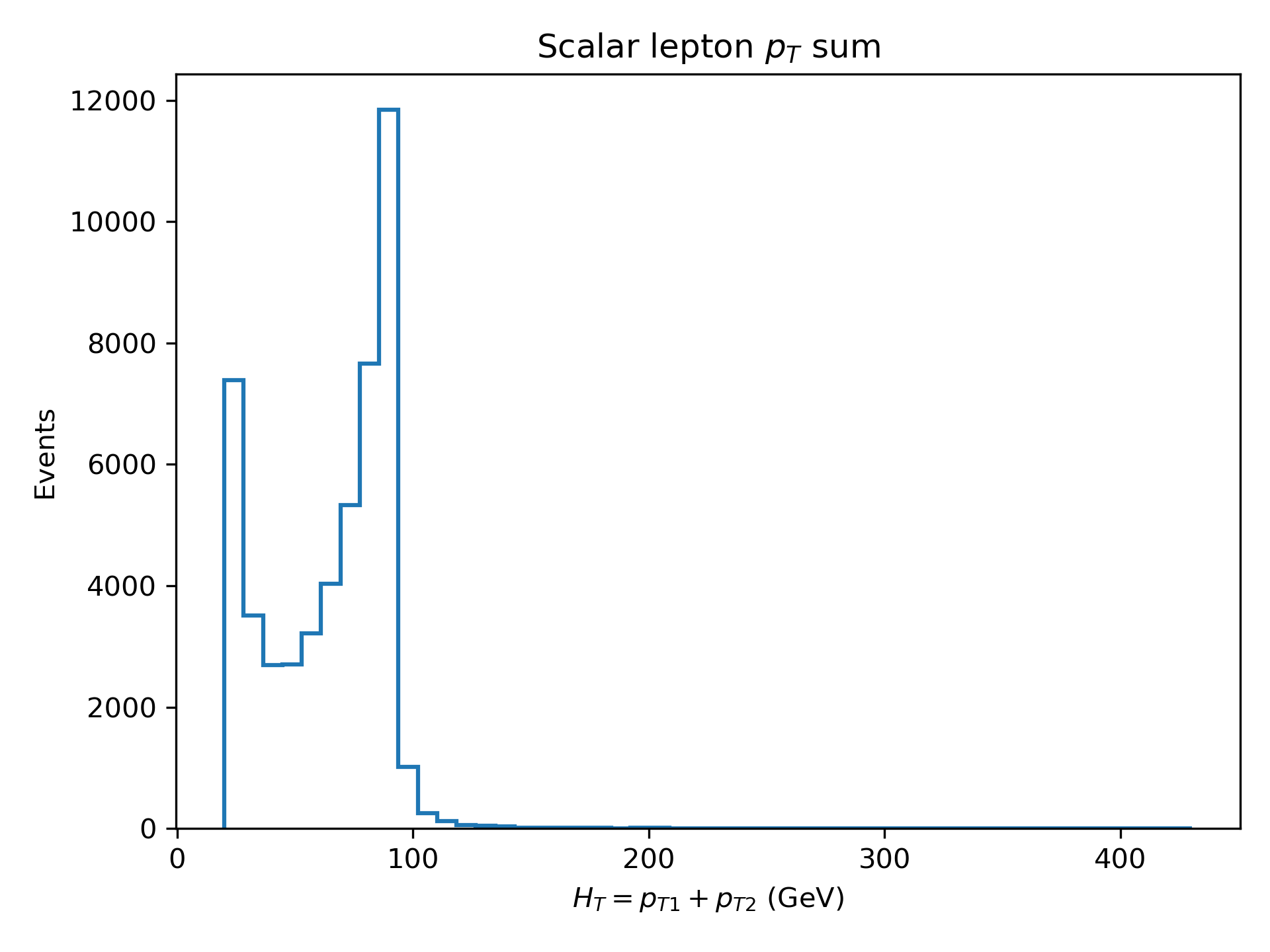} \ \ 
  \includegraphics[width=5.9cm]{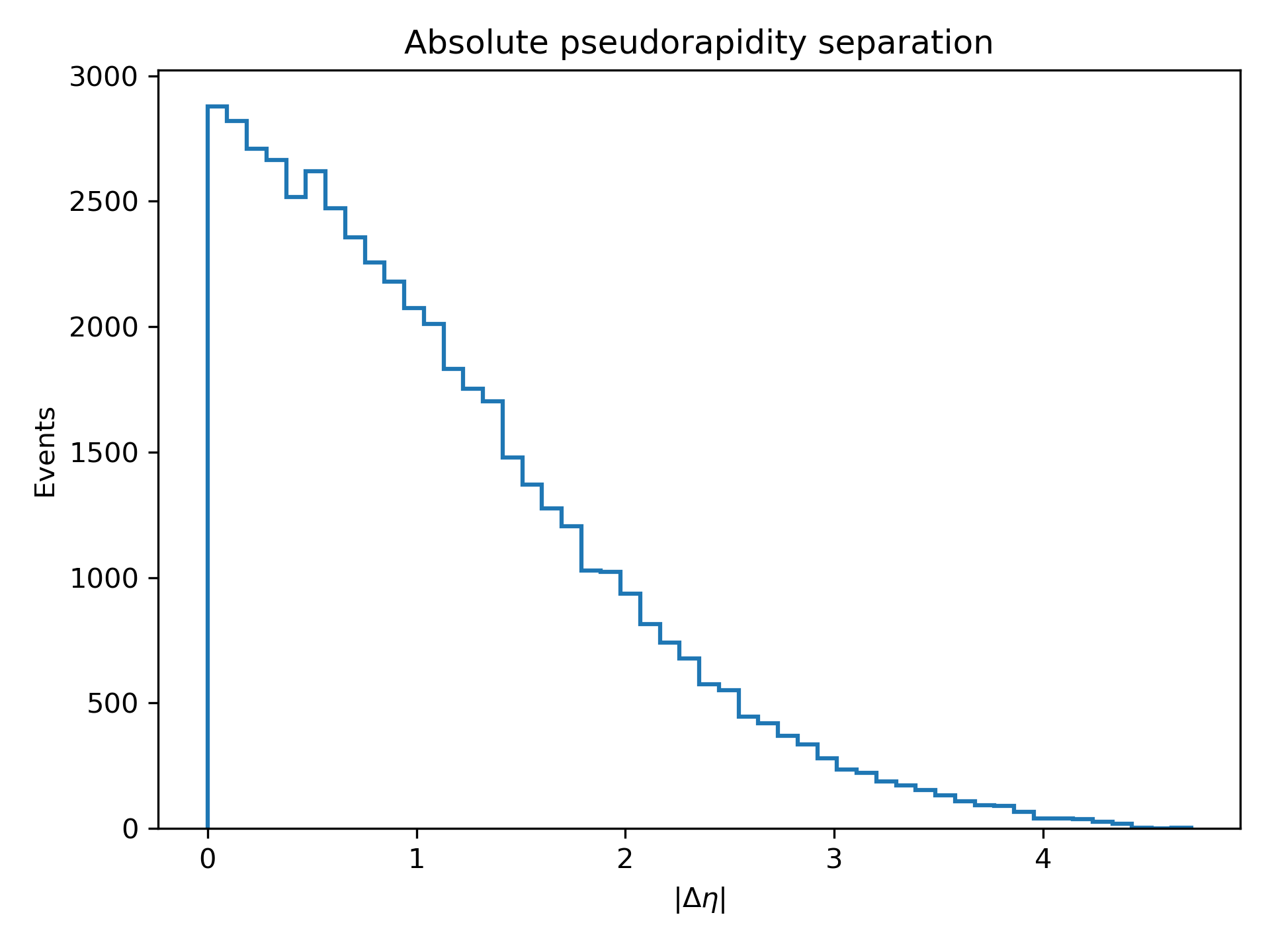} 
 \end{center}
\caption{\scriptsize Left : Distribution of the scalar sum of lepton transverse momenta $(H_T)$. This variable dictates the energy scale and is used as a primary coordinate in our regime partitioning.
\newline
Right : Distribution of the absolute pseudorapidity separation $|\Delta\eta|$ . High values of $|\Delta\eta|$ characterize forward-scattering events, which often suffer from higher detector noise.}
\label{fig:hteta}
\end{figure}

To describe the physical state of a system, the effectiveness of any fast reconstruction pipeline relies on the local assumptions. 
To test this assumption, this work uses $Z \to \ell \ell$ events generated by \textsc{MadGraph}5\_aMC@NLO \cite{Alwall:2014hca} as a benchmark in HEP. 
To focus on the $Z$ boson decay to the leptons, this study performs the parton level analysis avoiding the uncertainties introduced by the parton showering and the detector level smearing. 
By this approach, it is possible to examine the intrinsic behaviour of the reconstruction pipeline itself. 
Therefore, no external simulation environments result in the observed bias,  except the logic of the reconstruction pipeline. 
This justifies the sole use of \textsc{MadGraph}5\_aMC@NLO and the isolation of the use of \textsc{Pythia}8 as well as \textsc{Delphes}3. 

The generated datasets give the precise truth-level descriptions of the leptonic decay of the $Z$-boson. 
To provide a high-fidelity sample, we generate a dataset of $100000$ signal events, where each event is categorized by local parameters and global context variables. 
The local parameters are given by the individual four-momenta $p^{\mu}$ of the two leptons, and the global context parameters are given by the scalar sum $H_T$ of the two transverse momenta $p_{T_{1,\,2}}$ ($i.e.$, $H_T\, = \, p_{T_1} + p_{T_2}$) and the absolute pseudorapidity separation $|\Delta\eta|$. 

The unsupervised clustering identifies the natural groupings in the space of global contexts, whereas the supervised framework requires pre-labeled data, which restricts the scope of new discovery. 
Therefore, instead of supervised machine learning, this paper uses unsupervised machine learning to eliminate the constraints of the pre-labeled datasets along with the human bias. 
To group the global phase space of $H_T$ and $|\Delta\eta|$ into $k$ distinct context regimes, we use the K-Means clustering algorithm, which labels mathematically identical event environments without knowing the underlying physics. 
This step is domain-agnostic, as this clustering converts the continuous phase space into discrete regions. 
This facilitates the evaluation of the stability of the reconstruction pipeline across varied kinematic topologies. 

The K-Means algorithm inspects the global energy scale $H_T$ and the absolute value of the separation angle $|\Delta\eta|$ with equal mathematical weight. 
After generating the $100000$ points dataset \cite{Dillon:2020quc,Nachman:2021yvi}, we first shuffled that, and then splitted the entire dataset into three parts -- $70\%$ for the training of the regime map, $15\%$ for validation, and $15\%$ for the final stress testing. 

To estimate the robustness of the reconstruction pipeline, the reconstruction error ($\delta$) can be defined by,
\begin{equation}
 \delta = \frac{m_{reco} - m_{truth}}{m_{truth}}\,.
\label{eq:error}
\end{equation}
If one considers only the local assumptions, $\delta$ would remain constant irrespective of the global kinematic environment. 
If $\delta$ remains constant with varying global parameters, the reconstruction pipeline is robust; but if $\delta$ varies as a function of global parameters, the reconstruction pipeline is biased. 

Finally, to recover the truth-level physics, a lightweight correction layer is implemented, which takes the distinguished regime ID as the input feature.

\section{Results}
\label{Results}

As stated in the previous section \ref{Analysis}, this work uses \textsc{MadGraph5}\_aMC@NLO generated $100000$ points dataset for the signal $Z \to \ell \ell$ in HEP at $\sqrt{s}=14$ TeV. 
Some plots generated from this dataset clearly demonstrate the failure of the local assumption after a certain point. 
Also, the context-aware study of this paper exhibits a successful restoration of the resolution. 

The dilepton invariant mass distribution $m_{\ell\ell}$ over the mass ranges $60 - 120$ GeV is shown in the Fig. \ref{fig:mll}. 
The left plot shows the expected $Z$-boson resonance peak, and the right plot shows the dependence of $m_{\ell\ell}$ on the global context scale $H_T$. 
The plot exhibits that at lower energy scales, the reconstruction of mass is highly localized, but at higher energy scales, the points are significantly dispersed. 
This non-uniformity of the reconstruction resolution across the entire phase space suggests the requirement of unsupervised regime mapping. 

\begin{figure}
 \begin{center}
  \includegraphics[width=12cm]{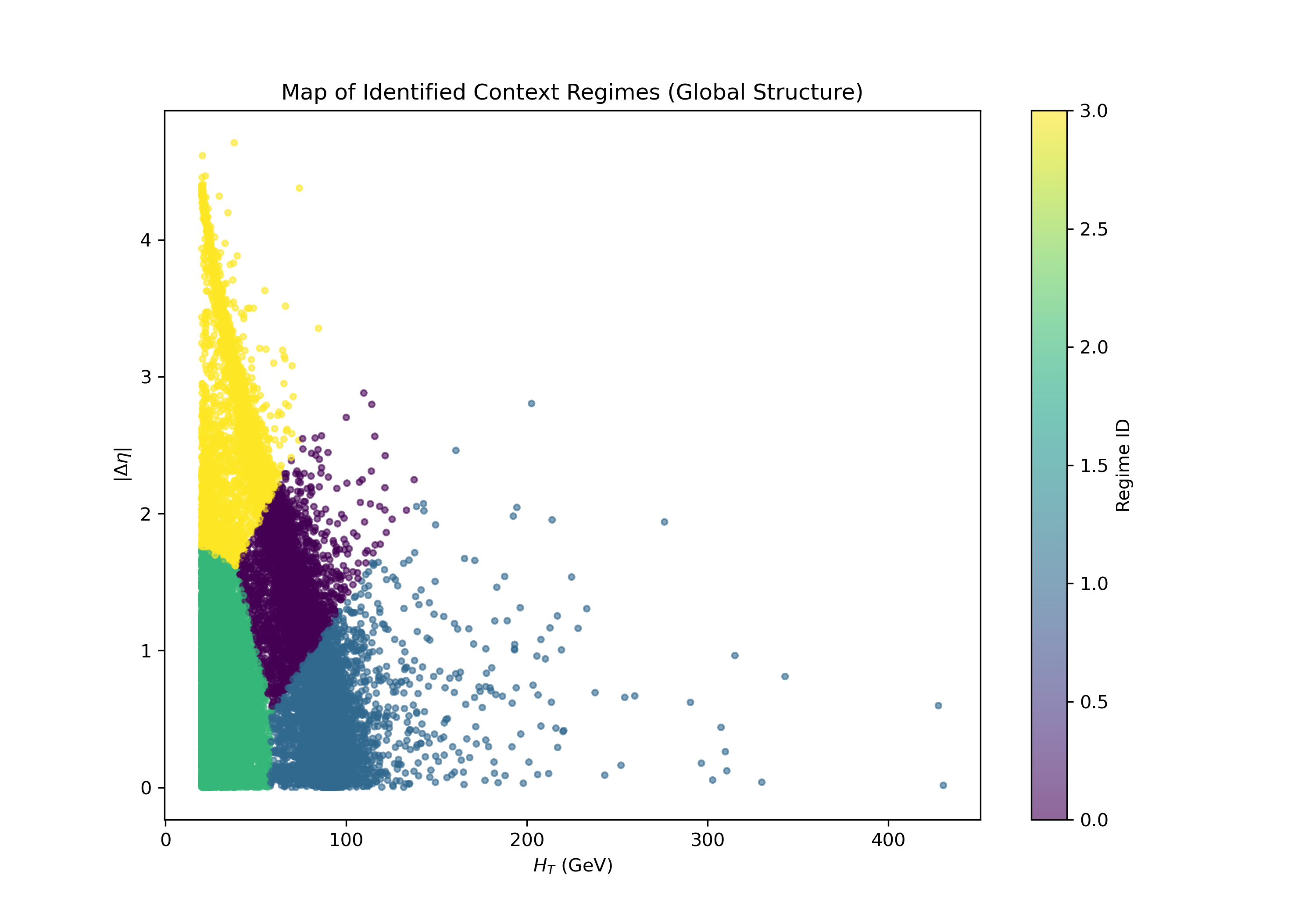} 
 \end{center}
\caption{\scriptsize Unsupervised clustering of the $(H_T , |\Delta\eta|)$ space into distinct context regimes.}
\label{fig:cluster}
\end{figure}

This shift in the $Z$-peak position is due to the influence of something outside the local parameters of the individual leptons, and we call this broader scenario as the ``global context''. 
We have two plots in the Fig. \ref{fig:hteta} -- the left plot shows the distribution of the scalar sum of lepton transverse momenta ($H_T$), and the right plot shows the distribution of their absolute angular separation ($|\Delta\eta|$). 
Traditionally, the reconstruction pipelines assume that the reconstructions are independent of the values of $H_T$ or $|\Delta\eta|$, but this work shows that high values of these parameters create a stress which cannot be guided properly by the local reconstructions. 

\begin{figure}
 \begin{center}
  \includegraphics[width=12cm]{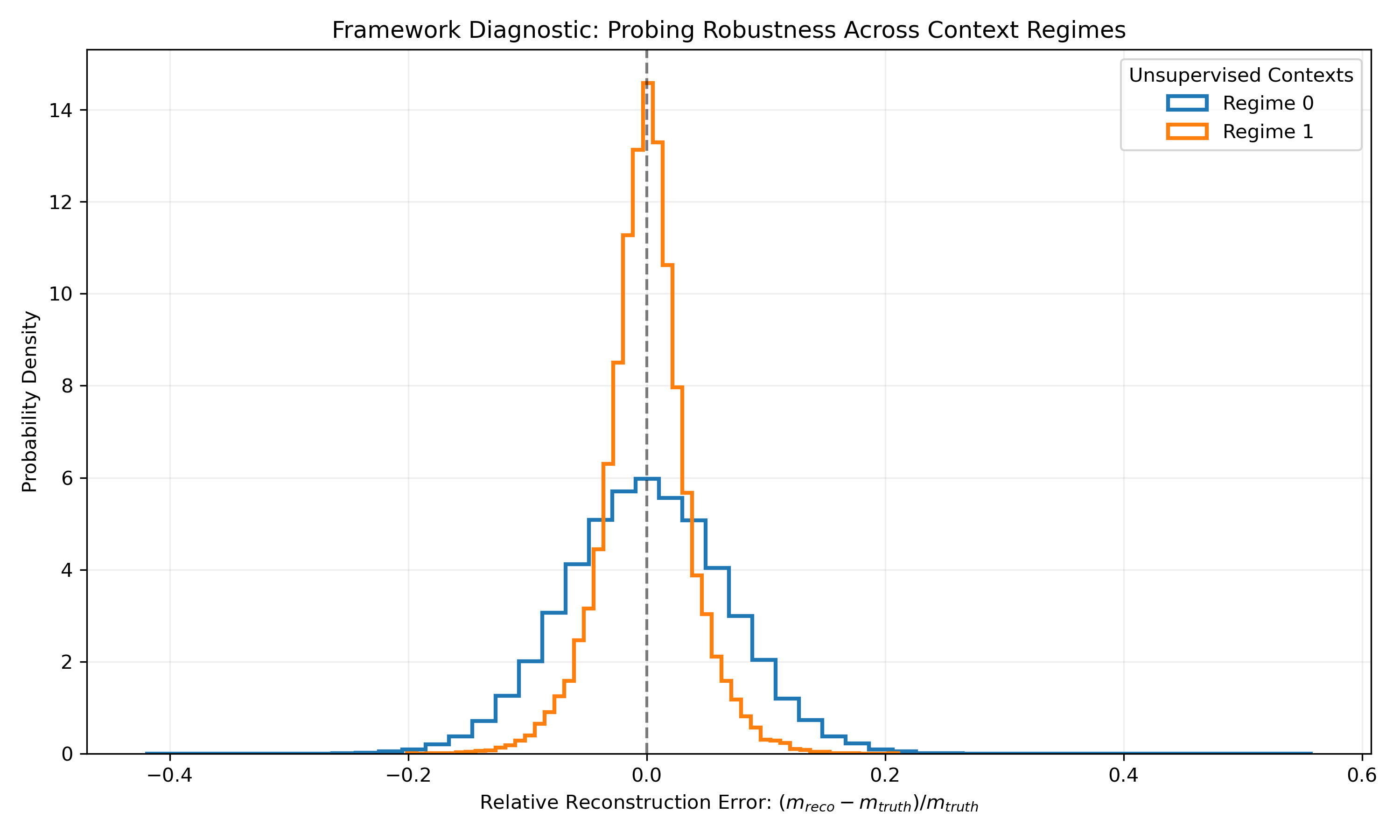} 
 \end{center}
\caption{\scriptsize Comparison of relative error distributions for different context regimes.}
\label{fig:error}
\end{figure}

The use of unsupervised machine learning organizes the data into different regimes as shown in the Fig. \ref{fig:cluster}. 
The framework automatically identifies distinct kinematic clusters based on the energy $H_T$ and the separation $|\Delta\eta|$, by the grouping of the events. 
The purple, blue, green, yellow points characterize regime $0,\,1,\,2,\,3$ respectively. 
The purple points represent moderate energy, medium separation events. 
The blue points denote high energy but low separation events. 
These two regions (regime $0\,\&\,1$) express the signal region for the $Z$-boson decays, where the leptons have moderate to high momenta and are relatively close to each other in terms of the separation angle. 
The green points stand for low values for energy and separation, both. 
This region (regime $2$) embodies the background region, where the leptons do not possess much energy and are not likely coming from the $Z$-boson decay. 
The yellow points (regime $3$) typify low energy and high separation events. 

The reconstruction algorithm has to perform harder for boosted particles, as small errors in angle or momentum measurement are amplified at high energies, and for this reason this work focuses only on the regime $0\,\&\,1$. 
Fig. \ref{fig:error} compares the error distribution $\delta$, as given in the Eqn. \ref{eq:error}, across these two regimes. 
If there were no errors ($\delta$=0), which is the ideal case, then we have the black dashed line. 
For regime $1$, the sharp orange peak is obtained, showing the robustness of this regime. 
Here, the distribution is narrow and tall and almost perfectly centered at $0$. 
This means the algorithm perfectly reconstructs the mass of the particles for the events in this regime. 
But the case is different for the regime $0$, where the broad blue curve is obtained, showing the non-robustness of this regime. 
Here, the distribution is wide, though still centered at $\delta\,=\,0$. 
This means the algorithm struggles here to reconstruct the mass of the particles for the events in this regime. 

\begin{figure}
 \begin{center}
  \includegraphics[width=12cm]{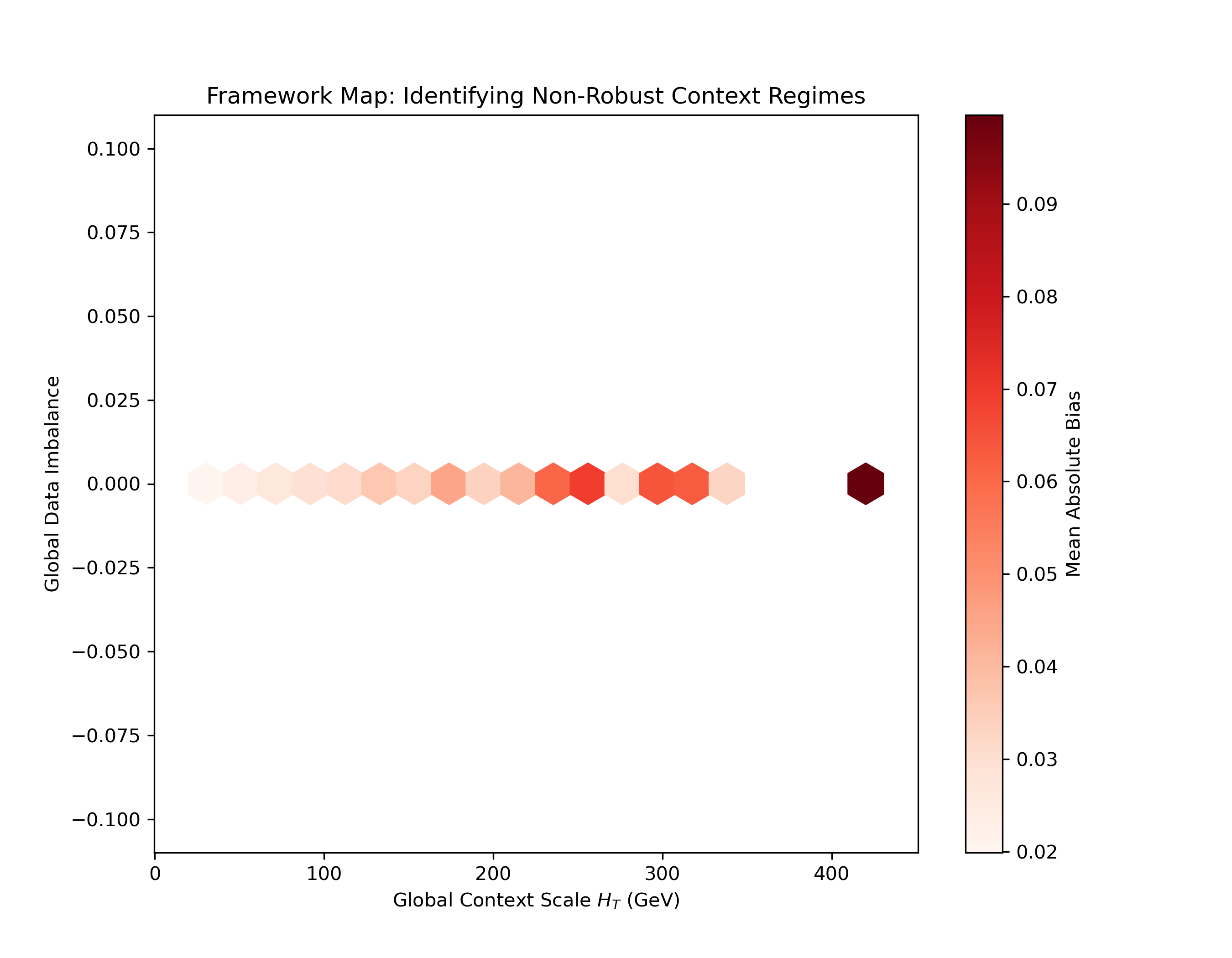} 
 \end{center}
\caption{\scriptsize Heatmap of mean absolute bias. Note the emergence of ``hot spots'' at high $H_T$ and high $|\Delta\eta|$, indicating systematic framework degradation.}
\label{fig:heatmap}
\end{figure}

The broadening of the distribution in the reconstruction error $\delta$ in the Fig. \ref{fig:error} hints at the interference of the global energy scale $H_T$ with the ability of the algorithm to reconstruct the particle's mass accurately. 
In this connection, Fig. \ref{fig:heatmap} identifies the non-robust context regimes as the function of the global context scale $H_T$, which is the scalar sum of the transverse momenta of the leptons. 
This heatmap tells us where the reconstruction pipeline becomes unreliable. 
A clear transition is seen here -- as we move from the low energy to high energy region, the mean absolute bias increases, which can be seen in the Fig. \ref{fig:heatmap} as the transition of light to dark shade of red, and we enter the danger zone ($>\,400\,$GeV) where the reconstruction pipeline starts to become erratic.

\section{Conclusion}
\label{Conclusion}

This work demonstrates that, the reconstruction pipeline in HEP is not globally uniform, rather it depends on the kinematic contexts, mainly on the reconstruction mass $m_{\ell\ell}$, scalar sum of the lepton transverse momenta $H_T$, and the absolute pseudorapidity separation $|\Delta\eta|$. 
The unsupervised clustering of data in the $H_T$ and $|\Delta\eta|$ feature space is grouped into different regimes, such that certain regimes provide stable and accurate mass reconstruction, and other regimes reveal systematic biases that necessitate further improvement. 
This regime-based approach provides a clear roadmap to improve framework robustness, shifting the focus from global average performance to targeted calibration within non-robust kinematic regions, which is a requirement of the HL-LHC era. 

\paragraph{A note on Medical Imaging Utility}
\label{Note} 
Though this work only includes HEP, the logic and strategies of this ``context-aware stress testing'' are directly applicable to the Medical Imaging (MI) reconstruction pipelines ($e.g.$ MRI or CT scan). 
In this regard, the scalar sum of transverse momenta $H_T$ and the invariant mass $m_{\ell\ell}$ in HEP would be replaced by Signal to Noise Ratio (SNR) and the tissue density, respectively, in MI. 
The identification of the non-robust regimes of the model in HEP resembles the minimization of the diagnostic misclassification in MI. 
We will extend our research in MI in this regard in future to explore the enhancement of the reliability of automated clinical imaging using unsupervised regime mapping. 

%
%

\end{document}